\documentclass[11pt]{article}
\usepackage{amsfonts, latexsym, amssymb,amsmath}
\newfont{\blackb}{msbm10 scaled\magstep1}
\def\Bbb#1{\hbox{\blackb #1}}
 
\begin{document}
\title{Special Functions of the Isomonodromy Type,
Rational Transformations of Spectral Parameter, and
Algebraic Solutions of the Sixth Painlev\'e Equation}
\author{A.~V.~Kitaev\footnote{E-mail: kitaev@pdmi.ras.ru} \\ 
\normalsize  
Steklov Mathematical Institute,\\
Fontanka 27, St. Petersburg, 191011, RUSSIA\\
and\\
Department of Pure Mathematics,\\    
University of Adelaide, \\
Adelaide 5005, AUSTRALIA} 
\date{May 20, 2000} 
\maketitle

\begin{abstract} 
We discuss relations which exist
between analytic functions belonging to the
recently introduced class of special functions of the
isomonodromy type (SFITs). These relations can be obtained 
by application of some simple transformations to 
auxiliary ODEs with respect to a spectral parameter which
associated with each SFIT. 
We consider two applications of rational transformations 
of the spectral parameter in the theory of SFITs.
One of the most striking applications which is considered
here is an explicit construction of algebraic solutions of
the sixth Painlev\'e equation. 
\end{abstract} 
\hspace{20pt}
\noindent
{\bf2000 Mathematics Subject Classification}: 34M55, 33E17

\section{Introduction} 
The general notion of special functions of the isomonodromy type (SFITs)
is introduced in the work \cite{K1}. It was shown there that
many classical special functions, e.g.,
the Gamma function, Gau{\ss} hypergeometric functions, Painlev\'e
functions, etc., belong to the class of SFITs. It was argued in 
\cite{K1} that a unique definition 
of such functions as the functions describing isomonodromy deformations 
of the matrix ODEs of the form,
\begin{equation}
\label{eq:ALAMBDA}
\frac{d\Psi}{d\lambda}=A(\lambda)\Psi,
\end{equation} 
where $\lambda$ is called throughout this paper the {\it spectral parameter} 
and $A(\lambda)$ is a $n\times n$ matrix-valued rational function of $\lambda$,
allows one to find many relations between these special functions, and
suggests a very natural and unique approach to study their properties by
using techniques developed for the matrix Riemann-Hilbert problem.
More precisely, define class of ODEs of the form
(\ref{eq:ALAMBDA}) under the following equivalence
\begin{equation}
\label{eq:eq12}
A\longrightarrow G^{-1}AG,\quad A\longrightarrow 
A+\partial_{\lambda}\log f(\lambda),
\end{equation}
where $G$ is $n\times n$ matrix independent of $\lambda$, 
and $f(\lambda)$ is an arbitrary scalar function of $\lambda$ with 
a rational logarithmic derivative.
We always suppose throughout the paper, that the function $f$ is chosen so that
$A(\lambda)\in sl_n(\Bbb C)$. 
Isomonodromy deformations are considered with respect
to {\it pole parameters}; to each pole $t_k^0$, $k=1,\ldots,m$, of order $p_k$ 
one can associate $(n-1)(p_k-1)$ "continuous" pole parameters, 
$t_k^{ij}$, $i=1,\ldots,p_k-1$, $j=1,\ldots,n-1$ via asymptotic expansion 
of the function 
$\Psi(\lambda)$ at $\lambda=t_k^0$ and $n-1$ "discrete" parameters 
$\theta_k^j$. With respect to the variables $t_k^{ij}$ and $t_k^0$
isomonodromy deformations of the coefficients of 
Eq.~(\ref{eq:ALAMBDA}), i.e., the corresponding SFITs, satisfy an 
overdetermined system of PDEs. With respect to the variables $\theta_k^{l}$ 
the SFIT solves a system of difference equations. One can write for SFITs
also differential-difference systems of equation
Due to the fractional-linear transformation of the spectral parameter 
$\lambda$, SFIT effectively depends on $(n-1)\sum_{k=1}^m(p_k-1)+m-3$ 
continuous variables and $m(n-1)$ discrete ones. 
Moreover, each SFIT depends on the monodromy 
variables, a point ${\cal M}$ of the so-called monodromy manifold
(an algebraic variety of the data characterizing monodromy group of 
Eq.~(\ref{eq:ALAMBDA})).
For those SFITs which solve linear equations,
monodromy manifold is just $\Bbb C^N$ for some natural 
$N$, and the dependence on the monodromy variables can be excluded by 
a specification of the linear independent solutions.
Instead of the set of the coefficients of Eq.~(\ref{eq:ALAMBDA}),
one may think of SFIT as about one function of the variables $t_k^{ij}$,
$t_k^0$, and $\theta_k^j$.
This can be done in a variety of ways, e.g., by means of the Jimbo-Miwa
$\tau$-function \cite{JM}.   

It is clear, that in this general setting, when we fix only the matrix 
dimension and functional dependence of $A(\lambda)$ as a rational function 
of $\lambda$, different equations of the type (\ref{eq:ALAMBDA}) 
define different SFITs, namely, they have either different number of 
continuous, discrete, or monodromy variables. Nevertheless, for special 
values of the variables, SFITs defined by Eq.~(\ref{eq:ALAMBDA}) with 
different $A(\lambda)$ may be related to each other by some explicit 
transformations. Many of these transformations can be found by a direct
group-theoretical analysis of the corresponding overdetermined systems 
of PDEs; however, due to the large number of variables, this analysis is 
quite complicated even with the help of a computer.  
An interesting question, therefore, is to consider transformations acting 
on the set of classes of equations of the type of Eq.~(\ref{eq:ALAMBDA}) 
which generate relations between corresponding SFITs. In Section \ref{sec2}, 
we consider some of these transformations; in particular, rational 
transformations of the spectral parameter are introduced.   

A combination of rational and Schlesinger transformations (we call them
$RS$-trans\-for\-mations) is a simple but powerful method which allows one 
to get many non-trivial results in the theory of SFITs. 
In Section \ref{sec3}, on the example of the sixth Painlev\'e equation 
($P_6$), we explain how one can construct algebraic SFITs.  
In Section \ref{sec4}, two particular examples of new
algebraic solutions of $P_6$ are constructed by this method.
In Section \ref{sec5}, another application of $RS$- transformations 
for the sixth Painlev\'e equation is considered. Namely, it is shown that 
there are some special points, in the complex plane of independent variable
of $P_6$, which are different from $0$, $1$, and $\infty$ and have the 
following property:
for each point there are, at least, two transcendental solutions of 
$P_6$ (with coefficients satisfying some simple restriction) such that
at this point the first solution has the pole, 
the second has the zero, and the monodromy data of both solutions can be 
explicitly calculated in terms of their Laurent or Taylor expansions at 
this point, respectively.
These special transcendental solutions are similar to the so-called 
symmetric solutions of the the first and second Painlev\'e
equations \cite{K4}. The latter property allows one to solve the so-called
{\it connection problem} for these solutions in terms of their
expansions in the special points mentioned above.

Constructions which are based on the $RS$-transformations for the
multivariable SFITs will lead to the functions which are algebraic
with respect to a given subset of their continuous variables, whilst
the special points which are discussed in the previous paragraph will
be some special hypersurfaces in the space of the continuous variables.

It is also interesting to discuss relations between different SFITs from 
the point of view of transcendency of SFITs. In fact, from the point of view 
of transcendental functions existence of these relations means that the SFITs 
which can be expressed in terms of other SFITs don't define new 
transcendental functions. Recently, this question for the Painlev\'e equations
has been intensively discussed in the literature due to the approach 
developed by H.~Umemura \cite{U,W}. A key notion in the Umemura's approach is 
the notion of classical functions. Actually, SFITs in many respects are 
``not worse'' than Umemura's classical functions; say, properties of the 
Painlev\'e functions can be studied to a much greater extent than properties 
of the most classical functions. Moreover, the study of the transcendency
of the Painlev\'e equations in the Umemura's setting has shown that among
all classical functions only those which are SFITs solve Painlev\'e 
equations. From this point of view it would be interesting to have a
kind of generalized Umemura's theory which would include ``reducibility''
of the Painlev\'e transcendents.
Therefore, it would be natural to introduce a notion of transcendency of 
a SFIT (denote it $\tau$) with respect to a given set of SFITs 
$G=\{\tau_1,\ldots,\tau_r\}$, where each SFIT of the set $G$ depends on 
the same or fewer number of continuous variables 
than $\tau$. In this setting, the problem is to determine the set of discrete 
and monodromy variables of $\tau$ such that $\tau$ belongs to the 
$G$-extension of the differential field (more precisely its multivariable
generalization) of classical functions.

A particular case of the $RS$-transformations, namely, a combination of 
quadratic transformations of the spectral parameter with the Schlesinger 
transformations, has already been used in the study of the Painlev\'e 
equations. It is shown that, they allow to get quite non-trivial results: 
quadratic transformations for the sixth Painlev\'e equation \cite{K2}, 
a quadratic transformation between the third and fifth Painlev\'e equations 
(reducibility of the fifth Painlev\'e equation to the third one for the 
special values of the discrete parameters), and quadratic transformations 
between different Lax representations for the fourth and third Painlev\'e 
equations, which follows from the "gauge equivalence" between AKNS and KN 
hierarchies of soliton equations. The last fact can be reformulated as 
an example of relations between different SFITs which are discussed in the 
previous paragraph.
  
\section{Transformations of SFITs}
 \label{sec2}
As explained in the Introduction any transformation of
the solution $\Psi$ corresponding to some isomonodromy deformation of
Eq.~(\ref{eq:ALAMBDA}), which maps it into 
a function $\Phi$ solving an ODE of the same type, i.e., Eq.~(\ref{eq:ALAMBDA})
with $A(\lambda)\to\hat A(\lambda)$ where $\hat A(\lambda)$ is a 
rational function of $\lambda$ whose coefficients depend isomonodromically
on its pole parameters, generates transformations of the corresponding SFITs.  
Here, we list some transformations of this kind.
\begin{enumerate}
\item
P-Transformations. If $\Psi_k$ satisfies $\frac{d\Psi_k}{d\lambda}=
A_k(\lambda)\Psi_k$
for $k=1,2$, then $\Phi=\Psi_1\oplus\Psi_2$ solves
$$
\frac{d\Phi}{d\lambda}=\left(\begin{array}{cc}
A_1&\hat0\\
\hat0&A_2
\end{array}\right)\Phi,
$$
where $\hat0$ are the matrices each of whose elements are equal to zero;
\item
T-transformations. Let vector-functions $\psi_k(\lambda)$ 
solve $\frac{d\psi_k}{d\lambda}=A_k(\lambda)\psi_k$
for $k=1,2$, then vector $\psi=\psi_1\otimes\psi_2$ solves
$$
\frac{d\psi}{d\lambda}=\left(A_1\otimes I_2+I_1\otimes A_2\right)\psi,
$$
where $I_k$ are identical matrices of the sizes $A_k$. 
\item
$D_\kappa$-Transformations. If $\Psi$ is a solution of 
Eq.~(\ref{eq:ALAMBDA}) and $\kappa$ any parameter, say,
$\lambda$, $t_k^{lk}$, or $\theta_k^l$, then 
$\Phi=\frac{d\Psi}{d\kappa}\oplus\Psi$ solves
$$
\frac{d\Phi}{d\lambda}=\left(\begin{array}{cc}
A&\frac{dA}{d\kappa}\\
\hat0&A
\end{array}\right)\Phi.
$$
This transformation is easy to generalize for any vector 
parameter $\kappa$;
\item
L-Transformations. The Laplace transformation,
$$
\Psi=\int\limits_C\chi(\mu)e^{\lambda\mu}\,d\mu,\quad
\Phi=(\chi^{(N)},\ldots,\chi)^T,
$$
where $T$ means transposition and $(N)$ denotes $N$th derivative 
with respect to $\mu$; $N+1$ is the sum of orders of poles of matrix 
$A(\lambda)$. More information concerning L-transformations can be
found in \cite{BJL}\footnote{I would like to thank the referee
for this comment}; 
\item
S-Transformations. The Schlesinger transformations of
Eq.~(\ref{eq:ALAMBDA}). These transformations are generated by 
elementary transformations of the following form,
$$
\Phi=R(\nu)\Psi,\quad \nu=\left(\frac{\lambda-a}{\lambda-b}\right)^{1/n} 
\mathrm{or}\quad\nu=(\lambda-a)^{1/n},
$$
where $R(\nu)$ is a rational function of $\nu$ with poles at $\nu=0$ and 
$\infty$, and $a$ and $b$ are parameters which may coincide with the poles 
of $A(\lambda)$;
\item
R-Transformations. Rational transformations of
the spectral parameter $\lambda\rightarrow\mu$,
\begin{equation}
\label{eq:RATIONAL}
\lambda=R(\mu),\quad\Psi(\lambda)=\Phi(\mu),
\end{equation} 
evidently transforms Eq.~(\ref{eq:ALAMBDA}) into ODE with respect to $\mu$
of the same type,
\begin{equation}
\label{eq:BMU}
\frac{d\Phi}{d\mu}=B(\mu)\Phi,
\end{equation}
moreover, generators of the monodromy group of Eq.~(\ref{eq:BMU}) can be 
calculated as multiplications of the corresponding monodromy matrices and 
their inverses for Eq.~(\ref{eq:ALAMBDA}).
Thus, transformation (\ref{eq:RATIONAL}) preserve the isomonodromic property.
\end{enumerate}
In the next two sections we consider applications of R- and S-transformations 
to the theory of the sixth Painlev\'e equation.
\section{Construction of Algebraic SFITs.\\
The Sixth Painlev\'e Equation}
 \label{sec3}
In general SFITs are transcendental functions of several variables, 
but for some special values of the discrete and monodromy parameters, 
these functions can be algebraic functions of a subset of continuous 
variables. The idea which is presented in this section can be applied 
for construction of algebraic SFITs defined by Eq.~(\ref{eq:ALAMBDA}) 
of Fuchsian type. We explain this idea taking as an example the sixth 
Painlev\'e transcendent. This function is related with one of the simplest 
equation of the type (\ref{eq:ALAMBDA}) in $2\times2$ matrices. 
Constructions of algebraic SFITs related with the isomonodromy deformations 
of ODEs in matrix dimension higher than 2 can be obtained via
application of the transformations given in Section \ref{sec2}. 
A modification of this method which allows to get more general construction 
of algebraic SFITs defined by Fuchsian equations in the matrix dimension 
higher than $2$ will be given elsewhere.

Let us recall basic facts concerning definition of the sixth Painlev\'e 
equation as a SFIT \cite{JM}.
Consider $2\times2$ matrix Fuchsian ODE with four singular points,
\begin{equation}
 \label{eq:AP6}
\frac{d\Psi}{d\lambda}=\left(\frac{A_0}{\lambda}+\frac{A_1}{\lambda-1}
+\frac{A_t}{\lambda-t}\right)\Psi,
\end{equation}
where we suppose usual conditions, 
$A_k\in sl_2(\mathbb C)$ for $k=0,\,1,\,t,\,\infty$,
$A_\infty\equiv A_0+A_1+A_t=-\frac{\theta_\infty}2\sigma_3,\;
\theta_\infty\neq0$, 
which is in fact (excluding one exceptional solvable case \cite{M}) 
also a normalization rather than a condition on $A_k$.
Consider the system of Schlesinger equations,
\begin{equation}
 \label{eq:S}
\frac{dA_0}{dt}=\frac1t[A_t,\,A_0],\quad
\frac{dA_1}{dt}=\frac1{t-1}[A_t,\,A_1],\quad
\frac{dA_t}{dt}=[\frac1tA_0+\frac1{t-1}A_1,\,A_t].
\end{equation}  
This system is the compatibility condition of Eq.~(\ref{eq:AP6})
with 
\begin{equation}
 \label{eq:T}
\frac{d\Psi}{dt}=-\frac{A_t}{\lambda-t}\Psi.
\end{equation}
We call system (\ref{eq:S}) the Schlesinger deformations of Eq.~(\ref{eq:AP6})
Any solution of system (\ref{eq:S}) define an isomonodromy deformation
of Eq.~(\ref{eq:AP6}). The general solution (the set of all solutions)
of the system (\ref{eq:S}) we call special function of the isomonodromy type
and denote it as $SF_4^2$. This function depends on one continuous variable $t$
and four discrete variables $\theta_0,\,\theta_1,\,\theta_t,\,\theta_\infty$,
the latters are nothing but the eigenvalues of the matrices $2A_k$, 
$k=0,\,1,\,t,\,\infty$. It follows from Eq.~(\ref{eq:S}) that $\theta_k$ are
independent of $t$. Thus the complete notation for this function is
$SF_4^2(t;\,\theta_0,\theta_1,\theta_t,\theta_\infty)$.
As the function of variable $t$ $SF_4^2$ is known to be closely related
with the classical sixth Painlev\'e equation, $P_6$,
\begin{eqnarray}
  \label{eq:P6}
  &\frac{d^2y}{dt^2}=\frac 12\left(\frac 1y+\frac 1{y-1}+\frac 1{y-t}\right)
\left(\frac{dy}{dt}\right)^2-\left(\frac 1t+\frac 1{t-1}+\frac 1{y-t}\right)
\frac{dy}{dt}+&\nonumber\\
&\frac{y(y-1)(y-t)}{t^2(t-1)^2}\left(\alpha_6+\beta_6\frac t{y^2}+
\gamma_6\frac{t-1}{(y-1)^2}+\delta_6\frac{t(t-1)}{(y-t)^2}\right),&
\end{eqnarray}
where $\alpha_6,\,\beta_6,\,\gamma_6,\,\delta_6\in\Bbb C$ are parameters.  
We need
explicit relation $SF_4^2\longrightarrow P_6$. Suppose that a set of matrices
$\{A_k\}$ solves system (\ref{eq:S}) and denote $A_k^{ij}$ corresponding
matrix elements of $A_k$. Note, that due to the normalization
$$
A_0^{12}+A_1^{12}+A_t^{12}=A_0^{21}+A_1^{21}+A_t^{21}=0,
$$
therefore equations,
$$
\frac{A_0^{ik}}{y_{ik}}+\frac{A_1^{ik}}{y_{ik}-1}+\frac{A_t^{ik}}{y_{ik}-t}=0,
$$
for $\{ik\}=\{12\}$ and $\{ik\}=\{21\}$ have in general situation
($A_1^{ik}+tA_t^{ik}\neq0$) unique solutions $y_{ik}$. These functions
solve Eq.~(\ref{eq:P6}) with the following values of the parameters, 
\begin{eqnarray}
 \label{eq:12}
y_{12}(t):&&\alpha_6=\frac{(\theta_\infty-1)^2}2,\;\;
\beta_6=-\frac{\theta_0^2}2,
\;\;\gamma_6=\frac{\theta_1^2}2,\;\;\delta_6=\frac{1-\theta_t^2}2,\\
y_{21}(t):&&\alpha_6=\frac{(\theta_\infty+1)^2}2,\;\;
\beta_6=-\frac{\theta_0^2}2,
\;\;\gamma_6=\frac{\theta_1^2}2,\;\;\delta_6=\frac{1-\theta_t^2}2.
\label{eq:21}
\end{eqnarray}
Instead of Eq.~(\ref{eq:P6}) one can associate with $SF_4^2$ the
so called $\tau$-function, which plays a very important role in applications. 
This function \cite{JM} is defined via the function $\sigma$,
\begin{eqnarray*}
&&\sigma(t)=\mathrm{tr}\big(((t-1)A_0+tA_1)A_t\big)+t\kappa_1\kappa_2-
\frac12(\kappa_3\kappa_4+\kappa_1\kappa_2),
\end{eqnarray*}
where
$$
\kappa_1=\frac{\theta_t+\theta_\infty}2,\;\kappa_2=
\frac{\theta_t-\theta_\infty}2,\;
\kappa_3=-\frac{\theta_1+\theta_0}2,\;\kappa_4=\frac{\theta_1-\theta_0}2.
$$
The function $\sigma$ solves the following ODE,
$$
t^2(t-1)^2{\sigma''}^2\sigma'+\left(2\sigma'(t\sigma'-\sigma)-
{\sigma'}^2-\kappa_1\kappa_2
\kappa_3\kappa_4\right)^2=(\sigma'+\kappa_1^2)(\sigma'+\kappa_2^2)
(\sigma'+\kappa_3^2)(\sigma'+\kappa_4^2),
$$
where the prime is differentiation by $t$. The $\tau$-function is 
defined up to a multiplicative constant as the solution of the following ODE,
$$
t(t-1)\frac{d}{dt}\ln\tau=\sigma(t).
$$

Now, we are ready to explain our construction of algebraic solutions for $P_6$.
Consider the following matrix form of hypergeometric equation,
\begin{equation}
 \label{eq:PHI}
\frac{d\Phi}{d\mu}=\left(\frac{\hat A}{\mu}+\frac{\hat B}{\mu-1}\right)\Phi,
\end{equation}
where we, following \cite{J}, parameterize the matrices $A$ and $B$ 
by three complex numbers, $\alpha$, $\beta$, and $\delta$,
$$
\hat A=\left(\begin{array}{cc}
-\frac{(\alpha+\beta)(1-\delta)+2\alpha\beta}{2\beta-2\alpha}&
\frac{\beta(\beta+1-\delta)}{\beta-\alpha}\\
-\frac{\alpha(\alpha+1-\delta)}{\beta-\alpha}&
\frac{(\alpha+\beta)(1-\delta)+2\alpha\beta}{2\beta-2\alpha}
\end{array}\right),\quad
\hat B=-\hat A+\frac{\beta-\alpha}2\sigma_3.
$$
Explicit formula for the fundamental solution $\Phi$ in terms of 
the Gau{\ss} hypergeometric functions can be found in \cite{J}. 
Here we don't use it, however this formula is very important
in applications. Now, we make $R$-transformation, 
$\mu=P(\lambda)/Q(\lambda)$, $\Phi(\mu)=\Psi(\lambda)$, 
where polynomials $P(\mu)$ and $Q(\mu)$ have no common roots. 
Define the rank of R-transformation,
$$
\mathrm{rank}(R)=\mathrm{max}\{\deg(P),\,\deg(Q)\}.
$$ 
The number of singular points of the function $\Psi(\lambda)$ 
is not greater than $3\cdot\mathrm{rank}(R)$. 
The function $R$ depends on $\deg(P)+\deg(Q)+1$ parameters. 
These parameters can be used to reduce a number of singular points of $\Psi$, 
moreover, one of them should remain free to play the role of the deformation 
parameter $t$. If $\mathrm{rank}(R)\ge3$, then the number of the parameters 
cannot be chosen such that the function $\Psi(\lambda)$ has four singular 
points. In this case we can further specify some of the discrete parameters 
$\theta_k$ such that, after a set of four singular points is chosen, all extra 
"unwanted" singular points can be removed by Schlesinger transformation.
Construction of algebraic solutions by this method requires therefore 
classification of all rational functions for which this program can be 
fulfilled. A suitable parameter for this classification is
$\mathrm{rank}(R)$. At the moment all transformations of 
$\mathrm{rank}(R)\le4$ are classified \cite{K3}. Transformations with 
$\mathrm{rank}(R)\ge5$ are under classification.

Another method to construct algebraic solutions of the sixth Painlev\'e 
equation is to classify all cases when one-parameter families of solutions 
of the sixth Painlev\'e equations, which are known to be expressible in terms 
of logarithmic derivatives of the general solution of the hypergeometric 
equation, are algebraic due to the special choice of the latter solution.

Based on the results concerning algebraic solutions, which are known by 
the time this paper is written, it seems reasonable to make the following\\ 
{\bf Conjecture}.
The methods explained in the last two paragraphs 
allows one to construct all algebraic solutions of the sixth Painlev\'e 
equation, perhaps with the help of the certain transformations 
given in the Okamoto's work \cite{O}.   

\section{Examples of Algebraic Solutions of the Sixth Painlev\'e Equation}
 \label{sec4}
My interest in the construction of algebraic solutions of the sixth Painlev\'e
equation is related with the works of Hitchin \cite{H1, H2}, Umemura \cite{U},
and Dubrovin and Mazzocco \cite{DM}. Note, that one of the steps
in the Umemura's approach to the notion of transcendency of the Painlev\'e
functions is to classify algebraic solutions. The work \cite{DM} shows that
such classification in the case of the sixth Painlev\'e equation is much more
complicated than for other Painlev\'e equations.

Here we use notation introduced in the previous section, say, $\Psi(\lambda)$
is a fundamental solution of Eq. (\ref{eq:AP6}) and $\Phi(\mu)$ is the 
fundamental solutions of Eq.~(\ref{eq:PHI}).

It is convenient to use the following notation for R-transformations,
$$
R(a_1+\ldots+a_{n_1}|b_1+\ldots+b_{n_2}|c_1+\ldots+c_{n_3}),
$$
where $\{a_p\}_{p=1}^{n_1}$, $\{b_q\}_{q=1}^{n_2}$, 
and $\{c_r\}_{r=1}^{n_3}$ are the sets of integers denoting 
multiplicities of images of the points $\lambda=0,\;1$, and $\infty$ 
respectively. It is clear that
\begin{equation}
 \label{eq:notrat}
\mathrm{rank}(R)=\sum\limits_1^{n_1} a_p=\sum\limits_1^{n_2} b_q=
\sum\limits_1^{n_3} c_r.
\end{equation}
We denote compositions of $R$ and $S$ transformations as $RS_k(m)$;
this symbol stands as a general notation for transformations 
from equation with $m$ singular points into equation with 
$k$ singular points. For particular transformations of this kind
we use notation $RS_k(\ldots|\ldots|\ldots)$ where the space
inside the brackets is separated by the vertical lines on $m$ boxes.
Each box contains a partition of the $\mathrm{rank}(R)$ into the sum of 
integers as it is explained for Eq.~(\ref{eq:notrat}). 
We define $\mathrm{rank}(RS_k)=\mathrm{rank}(R)$.

Consider first a very simple example, $RS_4(3|2+1|1+1+1)$. It means that
$$
\mu=\frac{\rho(\lambda-a)^3}{\lambda(\lambda-1)},\quad 
\mu-1=\frac{\rho(\lambda-b)^2(\lambda-t)}{\lambda(\lambda-1)},
$$
where
$$
\rho=\frac{(2b-1)^3}{27b^2(b-1)^2},\quad
a=-\frac{b(b-2)}{2b-1},\quad
t=-\frac{b(b-2)^3}{(2b-1)^3}.
$$
Solution of the system (\ref{eq:AP6}), (\ref{eq:T}) reads,
$$
\Psi(\lambda)=\left(J_a\sqrt{\frac{\lambda-a}{\lambda-b}}+
J_b\sqrt{\frac{\lambda-b}{\lambda-a}}\right)\Phi(\mu),
$$
where 
$$
J_a=\frac{\left(\begin{array}{cc}
\beta&-\beta\\
\alpha&-\alpha
\end{array}\right)}{\beta-\alpha},\qquad
J_b=\frac{\left(\begin{array}{cc}
-\alpha&\beta\\
-\alpha&\beta
\end{array}\right)}{\beta-\alpha},
$$
and
$\Phi(\mu)$ solves solves Eq.~(\ref{eq:PHI}) with
$$
\delta=\frac23,\quad\alpha=\frac16-\beta.
$$
Note that
$$
J_a^2=J_a,\quad 
J_b^2=J_b,\quad
J_aJ_b=J_bJ_a=0,\quad
J_a+J_b=I.
$$
Eigenvalues of the matrices $2A_k$ in Eq.~(\ref{eq:AP6}), denoted 
$\pm\theta_k$, are
$$
\theta_t=\frac12,\quad
\theta_0=\theta_1=-\theta_\infty=2\beta-\frac16.
$$
Algebraic solutions of the sixth Painlev\'e equation (\ref{eq:12}) 
and (\ref{eq:21}) are as follows,
\begin{eqnarray}
 \label{eq:Y3}
y_{12}(t)&=&\frac{(b-2)^2(6\beta(b+1)+1-2b)}
{(2b-1)(6\beta(b+1)(b-2)-5+2b-2b^2)},\\
y_{21}(t)&=&\frac{(b-2)^2(6\beta(b+1)+b-2)}{(2b-1)(6\beta(b+1)(b-2)+7-b+b^2)}.
\end{eqnarray}
\begin{equation}
 \label{eq:SIGMA3}
\sigma(t)=\frac{2(12\beta-1)^2b^3(b-2)+(6\beta^2-\beta)(180b^2-132b+30)+
12b^2-10b-1}{72(2b-1)^3},
\end{equation}
\begin{equation}
 \label{eq:TAU3}
\tau(t)=C\frac{(2b-1)^{\frac1{12}}\big[b(b-1)\big]^
{\frac52(\beta-\frac1{12})^2-\frac1{2^5}}}
{\big[(b+1)(b-2)\big]^{\frac92(\beta-\frac1{12})^2+\frac1{3\cdot2^5}}},
\end{equation}
where $C$ is independent of $b$. In the derivation of formulae 
(\ref{eq:Y3}) -- (\ref{eq:TAU3})
it is supposed that $\beta\neq\frac1{12}$, nevertheless, only minor 
modifications are required for
$\beta=\frac1{12}$ ($\theta_\infty=0$), in particular, 
the functions (\ref{eq:Y3}) -- (\ref{eq:TAU3})
satisfy corresponding ODEs for all complex $\beta$. 

Another transformation we consider here is a bit more complicated,
$RS_4(2+1+1|3+1|2+2)$. In this case,
$$
\mu=\frac{\rho\lambda_1(\lambda_1-1)(\lambda_1-a)^2}
{(\lambda_1-b)^2(\lambda_1-c)^2},\quad
\mu-1=\frac{(\rho-1)(\lambda_1-d)^3(\lambda_1-e)}
{(\lambda_1-b)^2(\lambda_1-c)^2},
$$
where 
$\lambda_1=\frac{e\lambda}{\lambda+e-1}$, $a,\,c,\,e$ and $\rho$ are
the following functions of complex parameters $b$ and $d$:
\begin{eqnarray*}
a&=&-\frac{d(-d^2-4bd^2+8b^2d^2+6bd-12b^2d+3b^2)} 
{4d^3-8bd^3-3d^2+12bd^2-6bd+b^2},\\ 
c&=&\frac{(4bd-b-3d)d}{8bd^2-8bd+3b-4d^2+d},\\
e&=&-\frac{b^2(d-1)(4bd-b-3d)^2}
{9b^2d+b^4-12b^3d-27b^2d^2+d^3-10bd^3-16b^3d^3+24b^3d^2+24b^2d^3+6bd^2},\\
\rho&=&\frac{(4d^3-8bd^3-3d^2+12bd^2-6bd+b^2)^2}
{d(d-1)(8bd^2-8bd+3b-4d^2+d)^2}.
\end{eqnarray*}
Now we construct solution of the system (\ref{eq:AP6}), (\ref{eq:T}) with
the parameter
\begin{equation}
 \label{eq:tde}
t=-\frac{(b-1)^2d^2(4bd-d-3b)^2(-d^2-4bd^2+8b^2d^2+6bd-12b^2d+3b^2)}
{(2bd-b-d)^3(b-d)^3}
\end{equation} 
as follows,
$$
\Psi(\lambda)=D^{-1}S^{-1}(d,c)S^{-1}(d,b)\Phi(\mu),
$$
where: $\Phi(\mu)$ solves Eq.~(\ref{eq:PHI}) with
$$
\delta=\frac56+2\alpha,\quad\beta=\frac12+\alpha;
$$
$$
S(x,y)=J_+\sqrt{\frac{\lambda_1-y}{\lambda_1-x}}+
J_-\sqrt{\frac{\lambda_1-x}{\lambda_1-y}},\qquad
J_+=\left(\begin{array}{cc}
0&1\\
0&1
\end{array}\right),\quad
J_-=\left(\begin{array}{cc}
1&-1\\
0&0
\end{array}\right);
$$
$$
D=\left(\begin{array}{cc}
1+\det D&1\\
1&1
\end{array}\right),
$$
\begin{equation}
 \label{eq:D}
\det D=-\frac{2b(b-1)(b-d)(2bd-b-d)(4bd-3b-d)(4bd-b-3d)}
{\alpha(6\alpha-1)(8bd^2-8bd+3b-4d^2+d)(8b^2d^2-8b^2d-8bd^2+6bd+b^2+d^2)^2}.
\end{equation}
Note, that $J_{\pm}$ are the orthogonal projectors,
$$
J_+^2=J_+,\quad 
J_-^2=J_-,\quad
J_+J_-=J_-J_+=0,\quad
J_++J_-=I.
$$
Introducing parameter
$$
s=\frac{d-b}{d(b-1)}+1,
$$
one rewrites Eq.~(\ref{eq:tde}) as
$$
t=\frac{(3s+1)^2(3s^2+6s-1)}{(s^2-1)^3}.
$$
Solutions (\ref{eq:12}) and (\ref{eq:21}) of the sixth Painlev\'e equation 
corresponding to the parameters:
\begin{equation}
 \label{eq:theta2}
\theta_\infty=\frac23,\quad\theta_0=\theta_1=\frac12\theta_t=2\alpha-\frac16,
\end{equation}
are as follows
$$
y_{12}(t)=\frac{(3s+1)(3s^2+6s-1)}{(s^2-1)(s^2+6s+1)},
$$
\begin{eqnarray}
 \label{eq:yy21}
y_{21}(t)&=&\frac{y_{12}(t)}{Q(s,\alpha)}(3+6\alpha+72\alpha s+
(228\alpha-42)s^2+(72\alpha-24)s^3+(6\alpha-1)s^4)\nonumber\\
&\times&(-2+3\alpha+(36\alpha-6)s+(114\alpha+2)s^2+(36\alpha+6)s^3
+3\alpha s^4),\\
Q(s,\alpha)&=&(31212\alpha^2-5202\alpha-2164)s^4+
(16848\alpha^2-2808\alpha-996)(s^5+s^3)\\
&+&(3960\alpha^2-660\alpha)(s^6+s^2)+
(432\alpha^2-72\alpha+36)(s^7+s)\nonumber\\
&+&(18\alpha^2-3\alpha-6)(s^8+1).\nonumber
\end{eqnarray}
Note, that solution $y_{12}(t)$ solves Eq.~(\ref{eq:P6}) for arbitrary
complex $\alpha$, whilst it does not depend on $\alpha$! This means that
$y_{12}(t)$ solves the following algebraic equation,
$$
\frac{t-1}{(y_{12}-1)^2}=\frac{t}{y_{12}^2}+\frac{4t(t-1)}{(y_{12}-t)^2}.
$$
The functions $\sigma(t)$ and $\tau(t)$ corresponding to the parameters
given by Eq.~(\ref{eq:theta2}) are as follows,
\begin{eqnarray*}
\sigma(t)\!\!&=&\!\!
\frac{s^2+6s+1}{72(s^2-1)^3}((1-36\alpha+216\alpha^2)(1+s^4)-6(s+s^3)
-(22-72\alpha+432\alpha^2)s^2),\\
\tau(t)\!\!&=&\!\!
C(s^2-1)^{\frac1{12}}\frac{\left((3s^2+6s-1)(s^2-6s-3)\right)^
{\frac{(12\alpha-1)^2}{24}-\frac1{18}}}
{\left(s(s+3)(3s+1)\right)^{\frac{(12\alpha-1)^2}{24}-\frac1{72}}}.
\end{eqnarray*}
We would like to note that whilst the construction of the function
$\Psi(\lambda)$ should be modified for some special values of
$\alpha$ (see, e.g. Eq.~(\ref{eq:D})), the formulae obtained
for the functions $y_{12}(t)$, $y_{21}(t)$, $\sigma(t)$ and $\tau(t)$
remain valid for all $\alpha\in\Bbb C$. Since the formula (\ref{eq:yy21})
is quite complicated, we write below the function $y_{21}(t)$ for some
particular values of $\alpha$:
$$
\alpha=0\;\mathrm{or}\;\alpha=\frac16,\; 
y_{21}(t)=\frac{(3s+1)^2(3s^2+6s-1)(s^4+24s^3+42s^2-3)}
{(s^2+6s+1)(1082s^4+498(s^5+s^3)-18(s^7+s)+3(s^8+1))},
$$
\begin{eqnarray*}
\alpha=-\frac1{12},&&
y_{21}(t)=-\frac{(3s+1)(s^2+18s+5)}{5(s^2+6s+1)(s^2-1)},\\
\alpha=\phantom{-}\frac1{12},&&
y_{21}(t)=-\frac{(3s+1)(3s^2+6s-1)(s^4+36s^3+46s^2-12s-7)^2}
{(s^2-1)(s^2+6s+1)Q(s,1/12)},
\end{eqnarray*}
$$
Q(s,1/12)=19046s^4+8904(s^5+s^3)+220(s^6+s^2)-264(s^7+s)+49(s^8+1).
$$
\section{$RS_4(3)$ of Rank $2$}
 \label{sec5} 
There are only two (modulo fractional-linear transformations 
$RS_3(3)$ and $RS_4(4)$ of rank $1$) transformations $RS_4(3)$ of rank $2$.
One of them generates an algebraic solution of the sixth 
Painlev\'e equation, whilst another gives a simplest example of 
the points on the complex $t$ plane, different from $0$, $1$, and $\infty$, 
for which there exist transcendental solutions of the sixth Painlev\'e 
equation whose monodromy data of the associated Eq.~(\ref{eq:AP6}) can be 
calculated explicitly in terms of their expansions at these points. 
More sophisticated examples (generated by $RS$-transformations of the 
ranks $3$ and $4$) of the points with such property are given in \cite{K3}.

It is noticed by many authors that $y(t)=\pm\sqrt{t}$ is a solution
of Eq.~(\ref{eq:P6}) provided the parameters satisfy the following relations:
$$
(\theta_\infty-1)^2=\theta_0^2,\qquad\theta_1^2=\theta_t^2.
$$ 
Here we show that this algebraic solution is generated by
transformation $RS_4(2|1+1|1+1)$; the first transformation which is
mentioned in the previous paragraph. Though this transformation does
not lead to any new solutions the construction given below can be
valuable for applications.
$$
\mu=\frac{(\lambda-\sqrt{t})^2}{(1-\sqrt{t})^2\lambda},\quad
\mu-1=\frac{(\lambda-1)(\lambda-t)}{(1-\sqrt{t})^2\lambda}.
$$
The solution of the system (\ref{eq:AP6}), (\ref{eq:T}) reads:
$$
\Psi(\lambda)=\left(J_0\sqrt{\frac{\lambda}{\lambda-\sqrt{t}}}+
J_{\sqrt{t}}\sqrt{\frac{\lambda-\sqrt{t}}{\lambda}}\right)\Phi(\mu),
$$
where $\Phi(\mu)$ solves Eq.~(\ref{eq:PHI}) with $\delta=1/2$ and
$$
J_0=\left(\begin{array}{cc}
0&0\\
-\alpha/\beta&1
\end{array}\right),\qquad
J_{\sqrt{t}}=\left(\begin{array}{cc}
1&0\\
\alpha/\beta&0
\end{array}\right)
$$
are the orthogonal projectors:
$$
J_0^2=J_0,\quad
J_{\sqrt{t}}^2=J_{\sqrt{t}},\quad
J_0J_{\sqrt{t}}=J_{\sqrt{t}}J_0=0,\quad
J_0+J_{\sqrt{t}}=I.
$$
Solutions (\ref{eq:12}) and (\ref{eq:21}) of Eq.~(\ref{eq:P6}) 
corresponding to the parameters:
\begin{equation}
 \label{eq:theta3}
\theta_\infty=\alpha-\beta,\quad\theta_0=\alpha-\beta-1,\quad
\theta_1=\theta_t=\alpha+\beta+\frac12,
\end{equation}
are as follows:
\begin{eqnarray*}
y_{12}(t)&=&-\sqrt{t},\\
y_{21}(t)&=&-\sqrt{t}\,
\frac{\alpha(1+2\beta)(1+t)-2\sqrt{t}(\alpha^2+\beta^2+\beta)}
{\beta(1+2\alpha)(1+t)-2\sqrt{t}(\alpha^2+\beta^2+\alpha)}.
\end{eqnarray*}
The functions $\sigma(t)$ and $\tau(t)$ corresponding to the
parameters (\ref{eq:theta3}) read:
\begin{eqnarray*}
\sigma(t)&=&\frac12(\beta^2+\alpha^2+\beta+\frac14)-\alpha(1+2\beta)\sqrt{t}-
\frac{t}{16},\\
\tau(t)&=&Ct^{-\frac1{16}}\left(1-\frac1t\right)^
{\frac12(\alpha^2+\beta^2+\beta+\frac18)}
\left(\frac{1+\sqrt{t}}{1-\sqrt{t}}\right)^{\alpha(1+2\beta)}.
\end{eqnarray*}

The second transformation mentioned in the beginning of this section is
$RS_4(2|1+1|2)$. Actually, this transformation maps three points into 
four ones; therefore there is no need to use $S$-transformations, and it 
coincides with $R_4(2|1+1|2)$. The transformation reads,
$$
\Psi(\lambda)=\Phi(\mu),\qquad \mu=\lambda^2.
$$ 
The function $\Psi$ solves the following equation:
\begin{equation}
 \label{eq:point-1}
\frac{d\Psi}{d\lambda}=\left(\frac{2\hat A}{\lambda}+\frac{\hat B}{\lambda-1}+
\frac{\hat B}{\lambda+1}\right)\Psi,
\end{equation}
The monodromy data of Eq.~(\ref{eq:point-1}) can be calculated in terms
of the monodromy data of the hypergeometric equation (\ref{eq:PHI}), 
i.e., in terms of matrix elements of $\hat A$ and $\hat B$. The matrix elements
of Eq.~(\ref{eq:point-1}) can be viewed as the initial data  at $t=-1$
for the system of Schlesinger equations (\ref{eq:S}). In the general
case, $\theta_\infty\neq\pm1$, both solutions
of Eq.~(\ref{eq:P6}), $y_{12}(t)$ and $y_{21}(t)$, which corresponds
to this deformation, has a pole at $t=-1$. By using parameterization
of the Schlesinger equations in terms of the solutions of the 
sixth Painlev\'e equation given in \cite{JM} one can completely
determine corresponding Laurent expansions of  $y_{12}(t)$ and $y_{21}(t)$
at $t=-1$. Note, that the parameters of formal monodromy corresponding
to this deformation are
\begin{equation}
 \label{eq:coeffpoint}
\theta_\infty=\beta-\alpha\neq\pm1,\quad
\theta_0=1-\delta,\quad
\theta_1=\theta_t=\beta+\alpha+1-\delta.
\end{equation}
Now, using transformation for the solutions of Eq.~(\ref{eq:P6}):
\begin{equation}
 \label{eq:involution}
y(t)=1/\hat y(\hat t),\;\;
t=1/\hat t,\qquad
\alpha_6=-\hat\beta_6,\;\;
\beta_6=-\hat\alpha_6,\;\;
\gamma_6=\hat\gamma_6,\;\;
\delta_6=\hat\delta_6,
\end{equation}
we get two solutions of Eq.~(\ref{eq:P6}), which are holomorphic at
$t=-1$, actually $\hat y_{12}(-1)=\hat y_{21}(-1)=0$, and due to the 
explicit formula (\ref{eq:involution})
relating them with solutions $y_{12}(t)$ and $y_{21}(t)$ we can by means
of the work \cite{J} find asymptotics of $\hat y_{12}(\hat t)$ as 
$\hat t\to0$, $1$, and $\infty$. It is also easy to find an
action of the involution (\ref{eq:involution}) on the $\Psi$-function.
Therefore, one can find explicitly initial (and monodromy) data of 
the Schlesinger system corresponding to both solutions 
$\hat y_{12}(\hat t)$ and $\hat y_{21}(\hat t)$.
By fractional-linear transformation of $\lambda$ preserving
the set of points $\{0,\,1,\,\infty\}$ ($R_4(4)$ in our notation)
the point $-1$ can be mapped to the points $1/2$ and $2$. So, that
these points have the same property as $-1$, i.e., for each point
there exist at least two transcendental solutions of Eq.~(\ref{eq:P6}), 
with the coefficients defined by Eq.~(\ref{eq:coeffpoint}) 
with the poles and two solutions with the zeros at this point, 
such that their monodromy data can be calculated explicitly in terms
of their Laurent or, respectively, Taylor expansions at this point. 
The word transcendental here means that these solutions are neither 
algebraic nor classical (in the Umemura sense) functions.  

{\bf Acknowledgment}
The author is grateful to J. Harnad and A. Its for invitation
to the Workshop on Isomonodromic Deformations and Applications
in Physics (Montreal, May 1--6, 2000) where this work was presented. 
  
\end{document}